\documentstyle[12pt]{article}
\newdimen\paperwidth
\newdimen\paperlength
\newdimen\margin
 \newdimen\vmargin
\textwidth=\paperwidth
\advance \textwidth by -2\margin
\advance\hoffset by -1truein
\advance\hoffset by \margin \textheight=\paperlength
\advance \textheight by -2\vmargin
\advance\voffset by -1truein
\advance\voffset by \vmargin
\advance\textheight by
-2\baselineskip
\begin{document}

\begin{titlepage} \title{ {\bf On the Application of} \\
{\bf the Non Linear Sigma Model }\\
{\bf to Spin Chains and Spin Ladders} }

\vspace{2cm} \author{
{\bf Germ\'{a}n Sierra}\thanks{Based on a talk delivered at the
Summer School on "Strongly Correlated Magnetic and Superconducting 
Systems", held in Madrid, Spain, July 1996 } \\ \mbox{}
\\
{\em
 Instituto de Matem\'{a}ticas y F\'{\i}sica Fundamental }\\ {\em
C.S.I.C., Madrid, SPAIN  } } \vspace{5cm}
\date{} \maketitle
\def\baselinestretch{1.3}

\begin{abstract}

We review the non linear sigma model approach (NLSM)  to spin chains
and  spin ladders, presenting new results. 
The generalization of the Haldane's map to ladders
in the Hamiltonian approach, give rise 
to different values of the $\theta$ parameter depending on the spin S, 
the number of legs $n_{\ell}$ and the choice of blocks needed to built 
up the NLSM fields. For rectangular blocks we obtain $\theta = 0 $ 
or $2 \pi S$  depending on wether  $n_{\ell}$, is even or odd, 
while for diagonal blocks we obtain 
$\theta = 2 \pi S n_{\ell}$. Both results agree modulo $2 \pi$,
and yield the same prediction, namely that even ( resp. odd) ladders 
are gapped (resp. gapless).  For even leeged ladders we show that the spin gap
collapses exponentially with $n_{\ell}$ and we propose a finite size 
correction to the gap formula recently derived by Chakravarty using 
the 2+1 NSLM, which gives a good fit
of numerical results. We show the existence of a Haldane phase in the
two legged ladder using diagonal blocks and 
finally we consider the phase diagram of dimerized ladders.

\end{abstract}


\vskip3in
\end{titlepage}

\newpage
\def\baselinestretch{1.5} \noindent

\section*{Introduction}

There have been three major developments in the 80's and 90's
in Condensed Matter or more specifically in Strongly Correlated
Systems. In historical order they are

\begin{itemize}

\item Haldane's conjecture in 1d antiferromagnetic 
spin chains (1983) \cite{Hal, AffLH}

\item Discovery of high-$T_c$ superconductivity and antiferromagnetism
in doped and undoped cuprate compounds (1986) \cite{Bed}

\item Discovery of Ladder Materials (1987,91) \cite{Lad87, Lad91}

\end{itemize}

All these findings have in common various features: low dimensionality,
antiferromagnetism, importance of quantum fluctuations and the fact that
they constituted theoretical and 
experimental surprises. 
Moreover these topics can be studied using sigma model techniques, which
establish a methodological link between them.  
In this talk we shall be  mainly concerned with
spin chains and ladders.

The first surprise deals with the behaviour of 1d antiferromagnetic
Heisenberg spin chains (AFH) as a function of the spin.
The Hamiltonian describing the AFH model
is given by,

\begin{equation}
H_{\rm chain}= J \sum^L_{n=1} {\bf S}(n) {\bf S}(n+1)
\label{1}
\end{equation}

\noindent
where $J>0$ is the exchange coupling constant and 
${\bf S}(n)$ is a spin-S matrix acting  at the $n^{th}$ site
of the chain.
Haldane's conjecture concerning the spectrum and 
spin correlations of the Hamiltonian (\ref{1}) is given in 
table 1 ,

\begin{center}
\begin{tabular}{|c|c|c|}
\hline
$2S$ & Spectrum & Correlations  \\
\hline
\hline
even & gapped & exponential decay  \\
\hline
odd & gapless & algebraic decay  \\
\hline
\end{tabular}
\end{center}
\begin{center}
Table 1: Behaviour of spin chains
\end{center}

Nowadays there is sufficient theoretical, numerical and experimental
evidence to support this conjecture, which should rather be called a
"theorem" despite of the lack of a rigorous mathematical proof \cite{TH}. 

The third  surprise, in the historical order shown above, 
deals with the behaviour of spin ladders as function
of the number of legs $n_{\ell}$ \cite{DRS} 
( for a review and references on the subject see \cite{DR}). 
A ladder is an array of $n_{\ell}$ spin chains 
coupled as in figure 1. The Hamiltonian of the system
is given by,

\begin{eqnarray}
& H_{\rm ladder} = H_{\rm leg} + H_{\rm rung}& 
\nonumber \\
& H_{\rm leg}= J \sum^{n_{\ell}}_{a=1} \; 
\sum^L_{n=1} {\bf S}_a(n) {\bf S}_a(n+1) & 
\label{2} \\
& H_{\rm rung} = J' \sum^{n_{\ell}-1}_{a=1} {\bf S}_a(n) {\bf S}_{a+1}(n)&
\nonumber 
\end{eqnarray}

\noindent
where ${\bf S}_a(n)$ are spin-S matrices located in the $a^{\rm th}$
leg at the position $n=1, \dots, L$. We consider periodic boundary
conditions alongs the legs, i.e. ${\bf S}_a(n) = {\bf S}_a(n+L)$, and
open BC's along the rungs.
The intraleg coupling constant $J$ is positive but the interleg coupling
constant $J'$ can be either positive or negative. The qualitative  
behaviour of spin 1/2 ladders is given in table 2,

\begin{center}
\begin{tabular}{|c|c|c|}
\hline
$n_{\ell}$ & Spectrum & Correlations  \\
\hline
\hline
even & gapped & exponential decay  \\
\hline
odd & gapless & algebraic decay  \\
\hline
\end{tabular}
\end{center}
\begin{center}
Table 2: Behaviour of Spin 1/2 Ladders
\end{center}

This result holds for both signs of $J'$. 
The analogy between the integer/half-integer behaviour of spin chains
and the even/odd behaviour of spin 1/2 ladders is evident.
In the limit where $J'$ goes to minus infinity, 
the spin ladder Hamiltonian (\ref{2})
becomes equivalent to the spin chain Hamiltonian (\ref{1})
for an effective spin equal to $n_{\ell}/2$, and hence the ladder
behaviour given in table 2 follows from the behaviour of the spin
chains. What is not so obvious is that this behaviour
holds not just for strong ferromagnetic rung couplings 
but also for antiferromagnetic ones, regardless their magnitude.
As for the Haldane conjecture there is by now sufficient
evidence to support the "ladder conjecture" coming from
Quantum Monte Carlo, exact diagonalization, mean field theory,
experiments on ladder materials like VOPO or cuprates, bosonization,
finite size analysis, sigma model, etc \cite{DR}.

In this talk we shall give a unified description
of the behaviour of spin chains  and spin ladders 
utilizing the non linear sigma model (NLSM) \cite{Sene,Ger,Mora}.
Let us first recall 
the logic underlying Haldane conjecture. It is based in the 
following map\cite{Hal, AffLH},

\begin{equation}
{\rm Low } \; {\rm Energy} \; {\rm Modes} \; {\rm of}
\;{\rm the} \; {\rm Spin} \; {\rm Chain}
\longrightarrow \; {\rm Non} \; {\rm Linear} \;{\rm Sigma }\;{\rm Model}
\label{3}
\end{equation}

\noindent 
which is obtained in the semiclassical limit where
the spin S becomes very large, although (\ref{3})
can be derived on more general grounds 
based on symmetry arguments \cite{We}.
However these type of arguments miss the theta term in the action,
which plays a crucial role in determining the physics of the model.
 
Hence using the properties of the (NLSM) one derives 
those of the low energy spectrum
of the spin chain. 

To arrive at the results of table 2 for the ladders 
we shall  follow the same logic as for spin chains, that is,
we shall  construct a map

\begin{equation}
{\rm Low } \; {\rm Energy} \; {\rm Modes} \; {\rm of}
\;{\rm the} \; {\rm Spin} \; {\rm Ladder}
\longrightarrow \; {\rm Non} \; {\rm Linear} \;{\rm Sigma }\;{\rm Model}
\label{4}
\end{equation}

\noindent
and use the properties of the NLSM.
In this way the NLSM provides  
a unified and economic approach to both problems.
Actually, the 2d Antiferromagnetic Heisenberg
model, can also be mapped into the O(3) 
NLSM in 2+1 dimensions \cite{Hall}.
The RG analysis at finite temperature of the
later model made in \cite{CHN} 
has unravelled many of the properties
of the undoped cuprate compounds ( second surprise).
On the other hand the NLSM's that appear in the
r.h.s. of the  maps (\ref{3}) and (\ref{4})
are defined in 1+1 dimensions. This poses the problem of the crossover
between 1d and 2d Heisenberg systems through spin ladders \cite{Chak}.

\section*{The Non Linear Sigma Model: A Primer}

Before we construct the maps (\ref{3}) and (\ref{4}),
it will be convenient to review the basics of the O(3) NLSM.
This model was proposed as a  toy version of QCD, as can be seen by
enumerating its main  features: asymptotic freedom,  
dynamical  mass generation, instantons, skyrmions, 
integrability , etc (for a review see \cite{Frad,Auer}).  
For these and other  reasons the NLSM 
still attract the interest of many physicists and mathematicians. 
In the case of spin chains and ladders the NLSM becomes
not just a toy model but a reallistic model describing the low energy physics!.

The O(3) NLSM is a relativistic quantum field theory in 1+1
dimensions whose field
${\bf \Phi}$ is a three component vector living  on the 2d-sphere $S^2$,

\begin{equation}
{\bf \Phi}^2 = 1, 
\label{5}
\end{equation}

The euclidean action of the model is
given by,

\begin{equation}
S  = \int d^2 x \; \left[  - \frac{1}{2 g}
\left( \partial_{\mu} {\bf \Phi}\right)^2 + {\rm i} \frac{ \theta}{ 8 \pi}
\; \epsilon_{\mu \nu}\;  {\bf \Phi}\cdot \left( \partial_{\mu} {\bf \Phi}
\times \partial_{\nu} {\bf \Phi}\right)  \right]
\label{6}
\end{equation}

\noindent
where $g > 0$ is the sigma model coupling constant, 
and $\epsilon_{\mu \nu}$ the 2d Levi-Civita symbol.
The quantity

\begin{equation}
W = \frac{ 1}{ 8 \pi} \int d^2 x
\epsilon_{\mu \nu}  {\bf \Phi}\cdot \left( \partial_{\mu} {\bf \Phi}
\times \partial_{\nu} {\bf \Phi}\right)   \in {\cal Z}
\label{7}
\end{equation}

\noindent
takes an integer value
for field configurations ${\bf \Phi({\bf x})} $  which
go to a fixed value, say ${\bf \Phi}_0$, 
when  $|{\bf x}| \rightarrow \infty$ ( this condition is required to 
have a finite action).
Compactifying the space-time into the sphere $S^2$, the integral
(\ref{7}) 
gives the winding number of the map $S^2$ (space-time) $\rightarrow
S^2$ (target space). The
parameter $\theta$ enters in the partition function as 
$ e^{ {\rm i} \theta W}$, and therefore the integer character of W 
implies  that $\theta$ is defined modulo $2 \pi$.
The theta term in the action is a truly topological 
term, which leads to dramatic non perturbative effects.

Notice that W  changes its sign under a parity 
( or time reversal)
transformations. Hence
the topological term of the action breaks explicitely 
parity (or time reversal)
unless $\theta = 0$ or $\pi$. Indeed if $\theta =0$ (mod 2 $\pi)$
the topological term is completely absent while if $\theta = \pi$
the winding number contributes to the action with a sign,
i.e. $ (-1)^W $. We expect from these properties that the AFH-spin chains
and ladders which are parity invariant will be associated with 
$\theta =0$ or $\pi$.

The basic properties of the NLSM for these
values of $\theta$ are given in table 3 \cite{ Polya,PW,HKS,Hal, Shan}.

\begin{center}
\begin{tabular}{|c|c|c|}
\hline
\hline
$ \theta $  ( mod  2$ \pi$)  & Spectrum & Correlations  \\
\hline
0 & gapped & exponential decay  \\
\hline
$\pi$ & gapless & algebraic decay  \\
\hline
\end{tabular}
\end{center}
\begin{center}
Table 3: Behaviour of the O(3) NLSM
\end{center}

This behaviour, which is independent of the magnitude of
the coupling constant $g$, will allow us to make a quick derivation
of table 3 in the strong coupling limit.
Before we do that, let's write the Hamiltonian that follows from
the action (\ref{6}),

\begin{equation}
H_{\rm NLSM } = \frac{c}{ 2} 
\int dx \; \left[ g \left( {\bf l} - \frac{ \theta}{4 \pi}
{\bf \Phi}'  \right)^2  +  \frac{1}{g}  {\bf \Phi'}^2 \right]
\label{8}
\end{equation}

\noindent
where  ${\bf \Phi}'(x) = \partial_x {\bf \Phi}(x)$,  $c$ is the ``speed
of light''
and ${\bf l}(x)$ is the 
angular momentum density, which is defined as,

\begin{equation}
{\bf  l} = {\bf \Phi} \times \frac{ d {\bf \Phi}}{ d t}
\label{9}
\end{equation}

Besides the constraint (\ref{5}) the fields ${\bf \Phi}$ and ${\bf l}$
satisfy,

\begin{equation}
{\bf l}(x) \cdot {\bf \Phi}(x) = 0
\label{10}
\end{equation}

\noindent
and the cannonical equal-time commutation relations,

\begin{eqnarray}
& [ l^a(x), l^b(y)] = {\rm i} \epsilon^{abc}
 \delta(x-y) l^c(x) & \nonumber \\
& [ l^a(x), \Phi^b(y)] = {\rm i} \epsilon^{abc}
\delta(x-y) \Phi^c(x) & \label{11} \\
& [ \Phi^a(x), \Phi^b(y)] = 0 &  \nonumber
\end{eqnarray}

It is very illustrative to prove the statements 
given in table 3 in the strong coupling limit 
($g >>1$ ), for which we shall use a regularized  lattice
version of the Hamiltonian (\ref{8}) \cite{HKS,Shan},

\begin{equation}
H_{NLSM}^{\rm lattice} =  
\frac{c}{ 2} 
\sum_n \; \left( g \; {\bf l}_n^2  
-  \frac{2}{g} \; {\bf \Phi}_n {\bf \Phi}_{n+1}  \right)
\label{12}
\end{equation}

At every site of the lattice there 
is a "particle" moving on a sphere 
(${\bf \Phi}^2_n =1)$, with angular momenta $l_n$.
If $\theta = 0$ the angular momenta
takes all possible integer 
values, $ l_n = 0,1, \dots \infty$ \cite{HKS}.
However if $\theta = \pi$ there is a monopole
of charge 1 at the center of every sphere, which 
implies that the 
possible values of the angular momentum are restricted to half-integers
values, $l_n = 1/2, 3/2, \dots \infty$ \cite{Shan}.

In the limit $g >>1$ the kinetic term $g \sum_n {\bf l}^2_n $ dominates
over the potential term $ - \frac{2}{g} \sum  {\bf \Phi}_n {\bf \Phi}_{n+1}$,
and  the ground state is obtained choosing the smallest possible value
of $l_n$ at every site. 

If $\theta =0$ the optimal choice is given by $l_n=0,\; \forall n$,
which yields a unique ground state. The first excited states are
obtained by choosing the irrep $l=1$ in one site and $l=0$ in
the rest of the chain. Since this can be done
at every site, there is a huge degeneracy, which is broken by the 
potential term, which delocalizes the $l=1$ excitations. 
The 3L degenerate first excited  states 
become a band of $l=1$  magnons, separated from the ground state by a 
gap, which can be computed in perturbation theory.
In \cite{HKS} this gap was computed up to $6^{th}$ order in $1/g^2$. 
The first three terms read,

\begin{equation}
\Delta = c  g \left( 1- \frac{2}{3} \frac{1}{g^2}+ 0.074 \frac{1}{g^4} 
+ O(\frac{1}{g^6}) \right) 
\label{13} 
\end{equation}

In the weak coupling limit ($g <<1$) 
a perturbative RG analysis shows that
the gap vanishes exponentially as \cite{Polya}

\begin{equation}
\Delta  \sim \frac{c}{g} e^{- 2 \pi/g}
\label{14}
\end{equation}

The proportionality constant depends on the regularization of the model
\cite{Shenker} (see also \cite{Hasen})
Combining the strong and weak coupling analysis 
and using Pade aproximants the authors of \cite{HKS}
concluded that the gap should never vanishes as long as
$g$ is non zero.

For $\theta = \pi$ the 
kinetic energy is minimized by the choice  
$l_n=1/2 \; \; \forall n$, which still leaves a huge
degeneracy for the ground state. This degeneracy is lifted by the
potential term, which leads to an effective AFH model when restricted to the
subspace $l_n=1/2$ \cite{Shan}. Since the $S=1/2$ AFH model 
is massless one gets that the NLSM at $\theta =\pi$ is also
massless. In reference \cite{Shan} it was argued that this gapless behaviour
persists to all values of $g$.

\section*{Haldane map for spin chains}

The map from the Heisenberg model into the NLSM can be done in various ways:
using coherent states in the path integral formalism, 
generalizing the Hubbard-Stratonovich formula in the partition function,
or applying gradient expansions in the Hamiltonian formalism. We shall
mainly use  the later one \cite{AffLH}, but we shall also briefly explain
the path integral approach for ladders.

The starting point of the construction is the spin wave analysis
of the AFH model. This consist in the linearization of
the evolution equations of the spin operators
${\bf S}(n)$, around the classical Neel configuration. 
The equations of motion
of the spin operators read,

\begin{eqnarray}
&\frac{ d {\bf S}(n)}{d t} = {\rm i} [ H_{\rm chain}, {\bf S}(n) ]&
\label{15} \\
& = -J {\bf S}(n) \times  [  {\bf S}(n+1) + {\bf S}(n-1) ] &
\nonumber 
\end{eqnarray}

The basic asumption is that ${\bf S}(n)$ deviates by a small  amount
from the alternating Neel configuration,

\begin{equation}
{\bf S}(n) = (-1)^n S \; {\bf z} + {\bf s}(n)
\label{16}
\end{equation}

\noindent
where $S$ is the magnitude of the spin and ${\bf z}$ is the unit 
vector in the vertical direction. In the linearized approximation
eq.(\ref{15}) becomes,

\begin{equation}
\frac{d \zeta(n)}{d t} = {\rm i} (-1)^{n+1} J S [ 
\zeta(n+1) + \zeta(n-1) + 2 \zeta(n) ]
\label{17}
\end{equation}

\noindent
where $ \zeta(n) = s^x(n) + {\rm i} s^y(n)$. 
The spin waves are the plane wave solutions of eq.(\ref{17}),

\begin{equation}
\zeta(n) = e^{ {\rm i} (\omega t + k n)} ( \psi(k) + (-1)^{n+1} 
{\bf \phi}(n) )
\label{18}
\end{equation}

We are interested in the low energy and 
long wavelength solutions  which are given
by,

\begin{eqnarray}
& \omega = v  k & \nonumber \\
& \psi(k) \sim A k & \label{19} \\
&  \phi(k) \sim B  & \nonumber 
\end{eqnarray}

\noindent
Equation (\ref{17}) is satisfied by the ansatz (\ref{18}) if  $2 A =B$. 
The spin wave velocity $v$ is given by,

\begin{equation}
v  = 2 J S
\label{20}
\end{equation}

The two  transverse spin wave solutions (\ref{18}) can be 
identified with the massless goldstone bosons 
associated with the 
breaking of the rotational  symmetry of the AFH Hamiltonian
by the classical Neel state. 
Eq(\ref{18}) implies that the spin variables have two slowly
varying components around the momenta $k=0$ ( field ${\bf l})$ 
and $k = \pi$ ( field ${\bf \Phi}$), which correspond to the 
local spin density and the local staggered magnetization respectively.

This elementary spin wave analysis is very useful in order to
construct the  map from
the Heisenberg model into the NLSM. This map can be mathematically
formulated as
a change of variables \cite{AffLH}. 
First of all, one has  to divide 
the chain into blocks of two sites ( see figure 2). 
The block, say (2n, 2n+1),
is given the coordinate x= 2n+1/2 and for every block one performs
the following change of variables,

\begin{eqnarray}
& {\bf S}(2n) = {\bf l}(x) - S {\bf \Phi}(x) & \label{21} \\
& {\bf S}(2n +1) = {\bf l}(x) + S {\bf \Phi}(x) & \nonumber 
\end{eqnarray}

The inverse of eqs.(\ref{21}) are

\begin{eqnarray}
& {\bf l}(x) = \frac{1}{2} [ {\bf S}(2n+1) + {\bf S}(2n)  ] &
\label{22} \\
& {{\bf \Phi}}(x) = \frac{1}{2S} [ {\bf S}(2n+1) -  {\bf S}(2n)  ] &
\nonumber
\end{eqnarray}

These relations imply that  ${\bf l}$ is a local
spin density and ${{\bf \Phi}}$ is a local staggered magnetization.
Using (\ref{21}) and the eqs ${\bf S}^2(n) = S(S+1)$ one gets,

\begin{eqnarray}
& {\bf l}^2(x) + S^2 {\bf \Phi}^2(x) = S( S+1) & \label{23} \\
& {\bf l} \cdot {\bf \Phi}=0 & \nonumber
\end{eqnarray}

In the semiclassical limit $S >>1$ eqs.(\ref{23}) 
become the sigma model conditions (\ref{5}) and (\ref{10}).
The commutators (\ref{11}) 
can also be obtained in the limit $S >>1$  from the commutation
relations of the spin matrices, 
with the following identification of the
Dirac's delta function and the Kronecker's delta symbol
in the limit where the lattice spacing $\delta$ goes to zero,

\begin{equation}
\delta(x-y) = \lim_{\delta \rightarrow 0} 
\frac{1}{2 \delta} \;\; \delta_{x,y}
\label{24}
\end{equation}

The term  $2 \delta$ in the denominator of 
eq.(\ref{24}) is  the lattice spacing in  the 
x-variables ( from now on we shall set $ \delta =1$).

The AFH Hamiltonian (\ref{1}) reads in terms of the
new variables ${\bf l}$ and ${\bf \Phi}$,

\begin{eqnarray}
&H_{\rm chain} = \sum_{x}  [- S(S+1) + 2 {\bf l}^2(x) 
+ {\bf l}(x) \; {\bf l}(x+2)& \label{25} \\
& + S ( {\bf \Phi}(x) \; {\bf l}(x+2) 
- {\bf l}(x)\; {\bf \Phi}(x+2) ) - S^2 {\bf \Phi}(x) \; {\bf \Phi}(x+2) ] &
\nonumber
\end{eqnarray}

Making  the asumption that the fields ${\bf l}(x)$ and
${\bf \Phi}(x)$ vary slowly in the $x$ coordinate, 
we can perform a gradient expansion and truncation of the
higher derivative terms. Keeping terms with at most two 
spatial or time derivatives in the field ${\bf \Phi}$,
we obtain the following Hamiltonian,

\begin{equation}
H_{\rm chain}^{\rm (Continuum)} = \int d x \; [ 2 {\bf l}^2 
- S ( {\bf l} {\bf \Phi}'+ {\bf \Phi}'{\bf l} ) + S^2 ({\bf \Phi}')^2 ]
\label{26}
\end{equation}

\noindent
which coincides with
the NLSM Hamiltonian given in (\ref{8}) upon the identifications,

\begin{equation}
\theta = 2 \pi S ,\;\;\; g = \frac{2}{S},\; \;\; c = 2 J S
\label{27}
\end{equation}

This is the desired result which permits to derive table 1 from table 3.
From eqs (\ref{14}) and (\ref{27}) we can estimate  the value of the spin  
gap and the correlation length as functions of the spin,

\begin{equation}
\Delta_S \sim J S^2 e^{-\pi S} , \; \; \xi \sim \frac{2}{S}
e^{\pi S}  
\label{28}
\end{equation}

In table  4  we show the  numerical 
values of the  gap, correlation length
and spin velocity of the  S= 1 and 2 Heisenberg chains,
obtained using Quantum Monte Carlo \cite{Taka,Yama} and
DMRG methods \cite{WH, Joli}.

\begin{center}
\begin{tabular}{|c|c|c|c|}
\hline
Spin &$\Delta/J$& $\xi$ & $c/J$ \\
\hline
\hline
$S=1$ & 0.4107 & $\sim 6$ & 2.47 \\
\hline
$S=2$ &0.049-0.085 & 49 & 4.16   \\
\hline
\end{tabular}
\end{center}
\begin{center}
Table 4: Spin gap of the $S=1$ and 2 chains.
\end{center}

This table shows the semi-quantitative agreement between the numerical
(exact) results and the NLSM estimates (\ref{28}). 
In the semiclassical limit where $S \rightarrow
\infty$ the NLSM predicts that the gap 
should go  to zero exponentially  fast, recovering in that
way  the classical gapless behaviour. This shows that the existence of
the gap is a truly quantum mechanical effect.

To end up our review of the spin chains, we would like to 
make some comments on the
non uniqueness of the map (\ref{3}). Indeed  eqs(\ref{21}) 
and (\ref{22}), which give
the map between the spin and the sigma model variables, 
depend on the choice
of the 2 site blocks. The other possible choice, given by the blocks
(2n+1, 2n+2), leads two another couple of variables $ {\tilde {\bf
l}}$ and ${\tilde {{\bf \Phi}}}$, which are  linearly related
to ${\bf l}$ and $ {\bf \Phi}$, by the eqs,

\begin{eqnarray}
& {\tilde {\bf l}} = {\bf l} - S \; {\bf \Phi}'& \label{29} \\
& {\tilde {\bf \Phi}}  = {\bf \Phi}&
\nonumber
\end{eqnarray}

These eqs. leave invariant
the constraints (\ref{5}), (\ref{10}), and in fact they 
can be obtained by the following  transformation \cite{AffLH},

\begin{eqnarray}
&  {\tilde {\bf l}} = \;
e^{- i S \int dx \;\beta' \; {\rm cos} \alpha } \; 
{\bf l} \; e^{ i S \int dx \;\beta' \; {\rm cos} \alpha }&
\label{30} \\
& {\tilde {\bf \Phi}} = \; 
e^{- i S \int dx \;\beta' \; {\rm cos} \alpha }\; 
{\bf \Phi} \; 
e^{ i S \int dx \;\beta' \; {\rm cos} \alpha }&
\nonumber 
\end{eqnarray}

\noindent
where $\alpha $ and $\beta$ are the spherical coordinates that 
parametrize the staggered field

\begin{equation}
{\bf \Phi} = ({\rm sin} \alpha 
\; {\rm cos} \beta,
{\rm sin} \alpha \; {\rm sin} \beta, {\rm cos} \alpha ) 
\label{alfa}
\end{equation}

However the Hamiltonian (\ref{26}) is not left invariant, but the 
change affects only  the theta parameter which for the new variables 
becomes,

\begin{equation}
{\tilde {\theta}} = - 2 \pi S
\label{31}
\end{equation}

This change in the value of $\theta$, depending on the blocking, 
can also be seen as  due to a parity transformation.
Recall that parity changes the sign of the topological term
W. The two blockings (2n, 2n +1) and (2n+1, 2n+2) 
are indeed related by parity.
In the next section we shall consider  more
general blockings which will lead to changes in $\theta$ but that are not
related by  parity transformations.
If $\theta = \pi$ the change (\ref{31}) leaves invariant the sign
factor $(-1)^W$ appearing in the partition function.

\section*{Haldane map for spin ladders}

There are two ways to study the problem of 
spin ladders using NLSM methods. The first one consists in the
application of the  Haldane's map  to every  
chain forming the ladder, obtaining a
system of $n_{\ell}$ sigma model  fields coupled by rung interactions.
This  approach is similar in spirit to the
bosonization studies of  spin ladders made in  references \cite{Bos}, and 
its applicability is reasonable in the weak coupling regime $J'/J <<1$.
The other possible approach is to built up a unique sigma model
field describing the low energy modes of the spin
ladder as a whole \cite{Sene,Ger,Mora}.
This approach should be  appropiate to  study  the 
intermediate coupling regime ($J'/J \sim O(1)$).
The  strong coupling regime ($ J'/J >>1$) has been mainly studied using
perturbative \cite{BDRS,RTR, BHSR}  
and mean field methods \cite{ GRS}. We shall see  that the later 
approaches are  related  to the discrete 
version of the NLSM based on the Hamiltonian (\ref{12}).

The map from the ladder into the NLSM follows the same steps as
that of the spin chain. We shall review below the approach of reference
\cite{Ger}.

\subsection*{Spin Wave Analysis of Ladders}

This analysis shows
that at the linearized level there are two massless modes 
corresponding  to two Goldstone bosons. 
These modes describe the gapless deviations 
from the  classical Neel solution, which 
in the case of AF-coupling along the rungs reads,

\begin{equation}
{\bf S}^{\rm class}_a(n) = (-1)^{a+n} S
\label{32}
\end{equation}

In the rest of the talk we shall confine ourselves to this case.
The transverse components of the spin waves 
, i.e. $\zeta_a(n) = s^x_a(n) + {\rm i} s^y_a(n)$ , are
given by

\begin{equation}
\zeta_a(n) = {\rm e}^{ {\rm i} ( \omega t + k n) }
\; ( A_a k + (-1)^{a+n +1} B )
\label{33}
\end{equation}

\noindent
where $\omega = v k$.  The quantities $A_a$ represent the fraction
of total spin carried out  by the $a^{th}-$leg, and
we shall normalize them as,

\begin{equation}
\sum_a A_a =1
\label{34}
\end{equation}

The spin wave velocity $v$ and  the $A_a's$ are given by,

\begin{eqnarray}
&\left( \frac{ v}{S} \right)^2 = J \; n_{\ell}/
\sum_{a,b} L^{-1}_{a,b} & \label{35} \\
& A_a =  \sum_b L^{-1}_{a,b} / \sum_{c,d} L^{-1}_{c,d} & \nonumber
\end{eqnarray}

\noindent
where $L^{-1}$ is the inverse of a matrix $L$ defined as follows,

\begin{equation}
L_{a,b} = \left\{ \begin{array}{cl} 
4 J + J' & a=b=1 \; {\rm or} \; n_{\ell} \\
4 J + 2 J' & 1< a=b < n_{\ell} \\
J' & |a-b| = 1 \end{array} \right.
\label{36}
\end{equation}

Besides the 2 Goldstone bosons there are, at the linearized level,
$2( n_{\ell} -1)$ massive modes. As an example 
we give below the values of their masses for the 
$n_{\ell}= 2$ and 3 ladders,

\begin{eqnarray}
& m_{n_{\ell}=2} = S \sqrt{8 J J'} & \label{37} \\
& m_{n_{\ell} =3} = S \sqrt{J' ( J+ 4 J')} \;\;, 
S \sqrt{ J' ( J + 12 J') } & \nonumber 
\end{eqnarray}

Of course in the limit when $J' \rightarrow 0$ the masses
of the massive modes  (\ref{37}) dissapear
and all the modes become massless, as should be the case for a set
of $n_{\ell}$ uncoupled chains. 
The  basic asumption in the construction of \cite{Ger} is the
truncation of the massive modes, keeping only
the gapless ones,  which is justified if the 
energy scales, generated non perturbatively,
are lower than the gap of the massive modes.
One may expect that this condition is satisfied in the intermediate
and strong coupling regimes.

\subsection*{ Map: Spin Ladders $\rightarrow NLSM$ }

The spin wave analysis serves as a  preparation for the more
complicated job of 
finding  the map from the
spin ladder into the sigma model. 
In the Hamiltonian formulation the first step 
is to split the ladder into blocks of $n_B$ sites, defining for each
block an average angular momentum ${\bf l}$ and staggered magnetization
${\bf \Phi}$ as follows,

\begin{eqnarray}
& {\bf l(x)} = \frac{n_{\ell}}{ n_B} \sum_{(a,n) \in B(x) } \;
{\bf S}_a(n) & \label{38} \\
& {\bf \Phi}(x)   = \frac{1}{ S n_B} \sum_{(a,n) \in B(x) } 
(-1)^{a+n}  \; {\bf S}_a(n) & \nonumber 
\end{eqnarray}

\noindent
where $x$ denotes the center of mass position of the block $B(x)$
along the leg axis, and the prefactor $n_{\ell}/n_B$ is required
in order that ${\bf l}$ satisfy the  cannonical 
commutation relations (\ref{11}) in the continuum limit. The  relation
(\ref{24}) reads in the more general case,

\begin{equation}
\delta(x-y) = \lim_{\delta \rightarrow 0}\;\; \frac{n_{\ell}}{n_B \delta} 
\delta_{x,y}
\label{39}
\end{equation}

For single chains there are only two types of blockings related by parity.
However for ladders there are 
many different types of blockings, which may
in principle lead to different results.  In order to choose those
blockings that are physically acceptable we shall   
impose the following  conditions

\begin{itemize}

\item  The blocks must have an even number of sites.

\item Every block must contain a site (or more ) of every leg the same
number of times, 
which implies that $n_B$ should be a multiple of $n_{\ell}$.

\item  Every site in a block must have a nearest neighbour belonging to
the same block.

\end{itemize}

Motivated by the
spin wave solution (\ref{33}), we shall propose
the following ansatz for the spin operators in terms of
the sigma model fields,

\begin{equation}
{\bf S}_a(n) = A_a \; {\bf l}(x) + (-1)^{a+n} S \; {\bf \Phi}(x)
\;\; {\rm for} \; (a,n) \in \;B(x) 
\label{40}
\end{equation}

\noindent 
The consistency between eqs.(\ref{38}) and (\ref{40}) is guaranteed by the
following identities,

\begin{eqnarray}
&\sum_{(a,n) \in B(x)}  (-1)^{a+n} = 0  & \nonumber \\
&\sum_{(a,n) \in B(x)}  A_a = \frac{n_B}{n_{\ell}}  & \label{41} \\
&\sum_{(a,n) \in B(x)} (-1)^{a+n}  A_a = 0  & \nonumber 
\end{eqnarray}

\noindent
which can be proved from the conditions on the allowed
blocks  given above and
the symmetry properties of $A_a$.

Strictely speaking we should add to the r.h.s. of (\ref{40})
a set of fields describing the massive modes whose 
existence we  discussed
in the spin wave analysis. Their inclusion makes the map 
(\ref{40}) more rigorous from a mathematical point of view,
allowing  a careful derivation of the effective
Hamiltonian governing the dynamics of the fields ${\bf l}$
and ${\bf \Phi}$ \cite{Ger}. Having found the effective Hamiltonian the
massive fields are discarded, so that we shall
not consider them from now on.

Let us now consider a few examples ladder's blockings.

\paragraph{ Columnar Blocks ( $n_B = n_{\ell}$) }

For ladders with an even numbers of legs the 
smallest possible blocks,
satisfying the  conditions given above,  coincide 
with the rungs (see Fig 3), i.e. $n_B = n_{\ell}$.
The partition of the ladders into rungs 
is what one effectively does in
the study of  the strong coupling 
limit of ladders \cite{BDRS,RTR,BHSR}.
In that sense one may expect to find relations between the NSLM
approach and the strong coupling analysis.

  Replacing  the ansatz (\ref{40}) into the ladder Hamiltonian (\ref{2}),
and taking the continuum limit we get the following expression,

\begin{equation}
H = \int dx \left[ \frac{1}{2} \;
L_{a,b} A_a A_b \; {\bf l}^2 + \frac{1}{2} J S^2 \; n_{\ell} \;  {\bf \Phi'}^2
\right]
\label{42}
\end{equation}

Comparing (\ref{42}) with the NLSM Hamiltonian (\ref{8}) we find,

\begin{equation} 
\begin{array}{cl}   \theta = & 0   \\
\left( \frac{c}{S} \right)^2 = & 
 J n_{\ell}/ \sum_{a,b} L^{-1}_{a,b} \\
g = &  1/ 
\left( S \sqrt{ J n_{\ell} \sum_{a,b} L^{-1}_{a,b} }\right)  \end{array}
\label{43}
\end{equation}

Since $\theta =0$ we see that the NLSM is gapped, which implies that
even ladders should always be gapped for any value of the spin.
The velocity $c$  coincides with the spin wave
velocity $v$ (\ref{35}). 
Finally, the coupling constant $g$ has a non trivial
dependence of the number of legs and the ratio $J'/J$. This is
interesting because we can derive from the eq.(\ref{43}) 
the dependence of the spin gap on the ladder's parameters.

For odd ladders the minimal blocks satisfying 
the conditions above  must have at least $n_B = 2 n_{\ell}$ sites. 
Unlike the 
blocks with $n_B = n_{\ell}$ there are now more geometries.
We shall in what follows consider the cases of even and odd
values of $n_{\ell}$. 
In figures 4 and 5 we show two of them. The choice of figure 4
was the one studied in \cite{Ger}. We shall give for 
completeness the results obtained there. Later on we shall study
the blocks of figure 5.

\subparagraph*{ Rectangular Blocks ($n_B = 2 n_{\ell}$) }

The Hamiltonian that one obtains 
after taking the continuum limit is (fig 4),

\begin{equation}
H = \int \frac{dx}{2} \left[ 
\sum_{a,b} L_{a,b} A_a A_b \;\; {\bf l}^2 + 2 J S^2 n_{\ell}\;{\bf \Phi}'^2
+ 2S J \sum_{a=1}^{n_{\ell}} (-1)^a A_a 
\left( {\bf \Phi}'\; {\bf l} + {\bf l} {\bf \Phi}'\right) 
\right]
\label{44}
\end{equation}

In \cite{Ger} it was shown that the value of $\theta$ corresponding
to (\ref{44}) is given by,

\begin{equation}
\theta = \left\{ \begin{array}{ll}
0 &  n_{\ell}: {\rm even} \\
2 \pi S & n_{\ell}: {\rm odd} 
\end{array} \right.
\label{45}
\end{equation}

The vanishing of $\theta$ for even ladders follows simply
from the symmetry property $ A_a = A_{n_{\ell}+1 -a}$,
while the  value $\theta=2 \pi S$
for odd ladders requires some more work \cite{Ger}.
The other NLSM parameters of (\ref{44}) are given by,

\begin{equation}
\begin{array}{ll}  \left( \frac{c}{S} \right)^2 = & 
2 \frac{ J n_{\ell} }{ \sum_{a,b} L^{-1}_{a,b} }
- \delta_{n_{\ell}, {\rm odd}} 
\frac{1}{ \left( 2 \sum _{a,b} L^{-1}_{a,b} 
\right)^2 } \\ 
 g = & 1/ 
\left( S \sqrt{2 J n_{\ell} \sum_{a,b} L^{-1}_{a,b}
- \frac{1}{4} \delta_{n_{\ell}, {\rm odd}}  }\right) \end{array}
\label{46}
\end{equation}

\noindent
where $\delta_{n_{\ell}, {\rm odd} } = 1$ ( resp. 0)  
if $n_{\ell} $ is odd ( resp. even).
For $n_{\ell}$ even the delta terms appearing in these eqs. are absent
and we recover (\ref{43}), 
except for a renormalization of $c$ and $g$
($ c_{\rm rect} = \sqrt{2} \;\; c_{\rm col} ,
\;\; g_{\rm rect} =  \;\; g_{\rm col}/ \sqrt{2} )$.

For $n_{\ell}$  odd, the formula (\ref{46}) does not coincide with 
the spin wave velocity (\ref{35}), except for the case of the
spin chains ( $n_{\ell} =1$)!. For large ladders the 
velocity approaches the value $4 J S$ ( $J'=J$) which  
differs by a factor of $\sqrt{2}$ with respect to the 2d value which is
$2\sqrt{2} JS$.

\subparagraph{ Diagonal Blocks $(n_B = 2 n_{\ell})$  }

Within the gradient approximation 
which we are using, it is consistent to
assign  to all the points within 
a diagonal block (fig. 5) the same coordinate $x$.
Under this asumption we get the following Hamiltonian,

\begin{equation}
H = \int \frac{dx}{2} \left[ 
\sum_{a,b} L_{a,b} A_a A_b \; {\bf l}^2 
+ 2 S^2 \sum_{a=1}^{n_{\ell}} p_a  {\bf \Phi}'^2
+ 2S \sum_{a=1}^{n_{\ell}} A_a p_a  
\left( {\bf \Phi}'\; {\bf l} + {\bf l} {\bf \Phi}'\right) 
\right]
\label{47}
\end{equation}

\noindent
where  $p_a$ is defined as

\begin{equation}
p_a = \left\{ \begin{array}{ll} J + J'/2 & a= 1 \; {\rm or} \; n_{\ell} \\
                                J + J' &  {\rm else} 
                                \end{array} \right.
\label{48}
\end{equation}

\noindent
The value of $\theta$ that comes out from (\ref{47}) is given by,

\begin{equation}
\theta = 8 \pi S \frac{ \sum_a A_a p_a}{ \sum_{b,c} L_{b,c} \; A_b A_c }=
8 \pi S \sum_{a,b} p_a L^{-1}_{a,b} 
\label{49}
\end{equation}

\noindent
To evaluate $\theta$ it is convenient to  write  (\ref{48}) 
in matrix form as follows,

\begin{equation}
\theta = 8 \pi S n_{\ell} <F| P L^{-1} |F>
\label{50}
\end{equation}

\noindent
where  $|F>$ denotes a normalized $n_{\ell}$-component 
vector with all its entries equal to $1/\sqrt{n_{\ell}}$ and
$P$  is a diagonal matrix with its entries given by $p_a$.
The trick is now to write  the matrix $L$ in the form 
$L = 4 P - K^-$, where $K^-$ is defined as,

\begin{equation}
K^-_{a,b}  = J' \times 
\left\{ \begin{array}{rl} 1 & a=b= 1 \; {\rm or}\; n_{\ell} \\
                                       2 & a=b \neq 1 \; 
{\rm or} \; n_{\ell} \\ -1  & |a-b| =1 \end{array} \right.
\label{51}
\end{equation}

\noindent
Making a "Dyson" decomposition of $L^{-1}$ 

\begin{equation}
L = \frac{1}{4P} - \frac{1}{4P} K^- \frac{1}{ 4 P - K^-}
\label{52}
\end{equation}

\noindent
and using the fact that $K^-$ annihilates the vector $|F>$ we finally
get,

\begin{equation}
\theta = 2 \pi S n_{\ell}
\label{53}
\end{equation}

It is quite remarkable that all the coupling constants in the
expression (\ref{49}) just combined in order to produce a result
which only depends on "global" data S and $n_{\ell}$. 
Eq.(\ref{53}) agrees with (\ref{45}) mod $2\pi$, 
implying that the choice of block does not affect this parameter.
Let us see what are the values of $g$ and $c$ for the Hamiltonian 
( \ref{47}),

\begin{equation}
\begin{array}{cl}  
\left( \frac{c}{S} \right)^2 = & 
2 \frac{ \sum_{a} p_a }{ \sum_{b,c} L^{-1}_{b,c} }- 
4 \left(  \frac{\sum_{a,b} p_a L^{-1}_{a,b} }{ \sum_{c,d} L_{c,d} }
\right)^2 \\
g= & 1/S \sqrt{ 2 \sum_{a,b,c} p_a L^{-1}_{b,c} - 
\left( 2 \sum_{a,b}  p_a L_{a,b}^{-1} \right)^2 } \end{array}
\label{54}
\end{equation}

\subsection*{Path Integral Derivation of Haldane Map for Ladders}

In this section we shall apply coherent state techniques to derive
the map from the spin ladders into the NLSM (for a review see  
reference \cite{Shankar2}). The map for the 2 legged ladder was
first worked out in \cite{Sene}. Our results agree
with the result of this reference and extend them to generic values
of $n_{\ell}$ ( see also \cite{Mora} for an alternative derivation
of the results given below).

The action of the spin ladder system in the  
path integral formulation can be written 
in the following form,

\begin{eqnarray}
& S =  \int d \tau \left\{ {\rm i}
\sum_{n} \sum_{a=1}^{n_{\ell}}   {\bf A}( {\bf S}_a(n) ) \cdot 
\frac{ d  {\bf S}_a(n) }{ d \tau } \right.  & \label{p1} \\
& \left. - J  \sum_{n} \sum_{a=1}^{n_{\ell}} {\bf S}_a(n) {\bf S}_a(n+1)
- J' \sum_{n} \sum_{a=1}^{ n_{\ell} -1} {\bf S}_a(n) 
{\bf S}_{a+1}(n) \right\} & \nonumber 
\end{eqnarray}

\noindent
where ${\bf S}_a(n)$ is a classical spin variable satisfying 
${\bf S}_a(n)^2 = S^2$ and ${\bf A}({\bf S}_a(n))$ is the vector
potential that fullfills the constraint ${\bf rot} {\bf A}({\bf n}) $
$= {\bf n} \;\; ( {\bf n} = {\bf S}/S )$.

The long wavelength limit of the ladder's spin variables 
$ {\bf S}_a(n) $  is given, according to (\ref{40}), by

\begin{equation}
{\bf S}_a(n) = \delta A_a {\bf l}(n) + (-1)^{a+n} S {\bf \Phi}(n)
\left( 1 - \frac{ \delta^2}{S^2} A^2_a {\bf l}^2 \right)^{1/2}
\label{p2}
\label{equation}
\end{equation}

\noindent
where the term with the square root is needed for the correct 
normalization of the variable ${\bf S}_a(n)$. 
For spin chains this later term can be replaced by 1, but 
for ladders it
gives a non trivial contribution when considering the
coupling between rungs.

Introducing (\ref{p2}) into (\ref{p1}) and performing the standard
gradient expansions one finds the following NLSM parameters

\begin{equation} 
\begin{array}{cl}   \theta = & 2 \pi S \; \delta_{n_{\ell}, {\rm odd}}   \\
\left( \frac{c}{S} \right)^2 = & 
 J n_{\ell}/ \sum_{a,b} L^{-1}_{a,b} \\
g = &  1/ 
\left( S \sqrt{ J n_{\ell} \sum_{a,b} L^{-1}_{a,b} }\right)  \end{array}
\label{p3}
\end{equation}

\noindent
which coincide with those obtained using columnar blocks
(\ref{43}) for even legged ladders. For odd ladders
the results of the path integral and those obtained
with rectangular blocks differ, 
except in the case of spin chains ($n_{\ell}=1$). 
This poses the problem between the relation between the Hamiltonian
formalism and the path integral formalism for this type of ladders.
In any case, notice that the path integral approach gives for even
and odd legged ladders a value of $c$ identical to the spin
wave velocity (\ref{35}).

Before we extract more consequences from the mapping of the ladders
into the NLSM, it is convenient to express $c$ and $g$ in terms of

\subsection*{ The function $f_{n_{\ell}}$}

Let us define $f_{n_{\ell}}$ as \cite{Ger},

\begin{eqnarray}
& f_{n_{\ell}}(J'/J)  = \frac{4 J}{ n_{\ell}} \sum_{a,b} L^{-1}_{a,b}&
\label{55} \\
& =\frac{1}{ n^2_{\ell} } \left[ \delta_{n_{\ell}, {\rm odd} }
+ 2 \sum_{m=1,3, \dots, n_{\ell}-1} \; 
{\rm sin}^{-2} \left( \frac{ \pi m}{ 2 n_{\ell} } \right)
\left( 1 + \frac{J'}{J} {\rm cos}^2 \frac{ \pi m}{ 2 n_{\ell} } 
\right)^{-1} \right]  &
\nonumber 
\end{eqnarray}

\noindent
Its explicit expression and numerical values for 
$z=1$ in the cases
$n_{\ell} =1, 2$ ,3, 4 and $\infty$ is given in table 5.

\begin{center}
\begin{tabular}{|c||c|c|c|c|c|}
\hline
$n_{\ell}$ & 1 & 2 & 3 & 4 & $\infty$ \\
\hline
$f_{n_{\ell}}(z) $ & 1 &$ 1/\left(1 + \frac{z}{2}\right)$
& $ \left( 1 + \frac{z}{12} \right)/\left( 1 + \frac{3 z}{4} \right) $& 
$ \left( 1 + \frac{z}{4} \right)/ \left(1 + z + \frac{z^2}{8}
\right)  $ & $\frac{1}{1+z}$     \\
\hline
$f_{n_{\ell}}(1)$ & 1 & 0.666 & 0.6190 & 0.5882 & 0.5 \\
\hline
$\frac{1}{2} \left(1+ \frac{1}{ n_{\ell}\sqrt{2}}  \right)$ 
& 0.853 & 0.677 & 0.6178 & 0.5884 & 0.5 \\
\hline
\end{tabular}
\end{center}
\begin{center}
Table 5   
\end{center}

$f_{n_{\ell}}(z)$ is a  monotonically decreasing
function in $z$,
which  varies from 1 to
$1/n^2_{\ell} \;\; \delta_{n_{\ell}, {\rm odd} }$ as $z$ varies from 0
to $\infty$. Interesting enough, the sum appearing in eq.(\ref{55}) can be
performed yielding,

\begin{equation}
f_{n_{\ell}}(z) = \frac{1}{1+z} \left[ 1+ \frac{z}{ n_{\ell} 
\left( 1 +z \right)^{1/2} } 
\frac{ \left( 1 + \frac{2}{z} \left( 1 + \sqrt{ 1+z} \right) 
\right)^{n_{\ell}} - (-1)^{n_{\ell}} } 
{ \left( 1 + \frac{2}{z} \left( 1 + \sqrt{ 1+z} \right) 
\right)^{n_{\ell}} + (-1)^{n_{\ell}} } \right]
\label{56}
\end{equation}

In  the isotropic case ($z=1$) we get,

\begin{equation}
f_{n_{\ell}}(1) = \frac{1}{2} \left( 1 + \frac{1}{n_{\ell} \sqrt{2} }
\frac{ \left( 3 + 2 \sqrt{2} \right)^{n_{\ell}} - (-1)^{n_{\ell}}  }{
\left( 3 + 2 \sqrt{2} \right)^{n_{\ell}} + (-1)^{n_{\ell}} } 
\right) 
\label{57}
\end{equation}

For $n_{\ell} > 2$, a good approximation of (\ref{57})
is given by ( see table 5),

\begin{equation}
f_{n_{\ell}}(1) \sim \frac{1}{2} \left( 1 + \frac{1}{n_{\ell} \sqrt{2} }
\right)
\label{58}
\end{equation}

In eqs (\ref{56}) and (\ref{57}) the difference between even and odd ladders
dissapears exponentially  as $n_{\ell}$ increases.

This ends the review of the properties of  $f_{n_{\ell}}$ and we return now
to our general discussion.

\subsubsection*{Summary of Results and Conclusions }

In table 6 we summarize the results obtained so far for the three
types of blocks we have considered so far,

\begin{center}
\begin{tabular}{|c|c|c|c|c|}
\hline
Block & $n_{\ell}$ & $\theta$ & $c$ & $g$   \\
\hline
\hline
Columnar & even & 0 & $2 J S / \sqrt{ f_{n_{\ell}} } $& 
$ 2/S n_{\ell} \sqrt{f_{n_{\ell}}} $     \\
\hline
Rectangular & even & 0 & $ 2 \sqrt{2} J S/\sqrt{f_{n_{\ell}}}$ &  
$\sqrt{2}/ S n_{\ell} \sqrt{f_{n_{\ell}}}$ \\
\hline
Rectangular & odd & $2 \pi S$ & 
$ \frac{ 2 \sqrt{2} J S}{ \sqrt{f_{n_{ \ell}}}}  \left( 1-
\frac{1}{ 2 n_{\ell}^2  f^2_{n_{\ell}}} \right)^{1/2}$ &
$ \frac{ \sqrt{2}}{  S n_{\ell} \sqrt{ f_{n_{\ell} } } }
\left( 1 - \frac{1}{2 n^2_{\ell} f_{n_{\ell}} } \right)^{-1/2}$ \\
\hline
Diagonal & both & $2 \pi S n_{\ell}$ &
$\frac{2 J S}{f_{n_{\ell}} } \left[ 2 f_{n_{\ell}} 
( 1 + (1- \frac{1}{n_{\ell}}) \frac{J'}{J} )- 1 \right]^{1/2}$  &
$ \frac{2}{S n_{\ell}}  \left[ 2 f_{n_{\ell}} 
( 1 + (1- \frac{1}{n_{\ell}}) \frac{J'}{J} )- 1 \right]^{-1/2}$  \\
\hline
Path Int. & both & 
$2 \pi S \delta_{n_{\ell},{\rm odd}}$ & $2 J S / \sqrt{ f_{n_{\ell}} } $& 
$ 2/S n_{\ell} \sqrt{f_{n_{\ell}}} $     \\
\hline
\end{tabular}
\end{center}
\begin{center}
Table 6
\end{center}

From this table we can extract the following  conclusions,

\begin{itemize}

\item The values of $\theta , c $ and $g$
are block dependent. For example, for $n_{\ell}$ even 
we get $g_{\rm rect} = \frac{1}{\sqrt{2}} 
\; g_{\rm col}$. Since the size of both blocks is different
we could interpreted the previous relation 
as the RG relation bewtween $g's$ at two different
length scales
($n_B({\rm rect})/n_B({\rm col}) = 2$). 
Curiously enough the value of $g_{\rm rec}$
is smaller than the value of 
$ g_{\rm col}$, suggesting that this "one step RG flow"  is
similar to the RG flow of the 2+1 NLSM, where the coupling constant
$g$ decreases at longer distances in the "renormalized
classical region" \cite{CHN}. After this first
RG step, which truncates the
ladder's degrees of freedom down  to those of the 1+1 NLSM,  
the value of $g$ will start growing as a consequence of the 
1+1 NLSM RG equations, so that the physics of the system will be
dominated by the strong coupling regime.

From table 6 we find two  "RG-invariant"
quantities: the $\theta$ parameter and the 
perpendicular spin susceptibility.

\item The invariance of the $\theta$ parameter
relies on its periodicity, which implies that 
for all purposes we may take $\theta =2 \pi S n_{\ell}$.
This is really a topological result for it only depends on the 
"global data" $S$ and $n_{\ell}$, and not on the values of the
ladder coupling constants $J$ and $J'$. For odd ladders this
result must be a consequence of the generalization, due to Affleck
\cite{Aff2}, of the well known   
theorem by Lieb-Schultz-Mattis 
\cite{LMS}, that asserts that in the infinite length limit the odd ladders
must either have a degenerate ground state or else there are gapless
excitations.

\item  The value of the bare perpendicular susceptibility $\chi^0 $ 
is given for all the block choices by,

\begin{equation}
\chi^0 = \frac{1}{ c \; g} = \frac{ n_{\ell} f_{n_{\ell}}}{4J} 
\label{59}
\end{equation}

\noindent

Notice that for large  $n_{\ell}$ 
the susceptibility per site 
$\chi^0/n_{\ell}$ goes to a finite value.

\end{itemize}

It is quite interesting to compare (\ref{59}) and the spin wave 
velocity (\ref{35}),

\begin{equation}
c = \frac{2 J S}{ \sqrt{ f_{n_{\ell}}}}
\label{60}
\end{equation}

\noindent
with the corresponding expressions of the bare 
perpendicular spin susceptibility and
spin wave velocities of a $d$ dimensional sigma model \cite{Manou},

\begin{equation}
\chi^0 = \frac{1}{ 4 d J a^d} \; , \;\; c= 2 \sqrt{d} J S a 
\label{61}
\end{equation}

\noindent
where $a$ denotes the lattice spacing.

The comparison of (\ref{59}), (\ref{60}) and (\ref{61}) suggest
somehow that ladders behave as  $d_{\rm ladder}$ dimensional  
spin systems with,

\begin{equation}
d_{\rm ladder} = \frac{1}{f_{n_{\ell}}}
\label{62}
\end{equation}

This naive definition of "fractal" dimension of ladders
helps to explain some numerical facts. First of all, if we choose
$J'=0$, then $f_{n_{\ell}}(0)=1$ and we get $d_{\rm ladder} = 1$, which
indeed corresponds to  1d chains. For the isotropic
models and  $n_{\ell}$ large we get,

\begin{equation}
d_{\rm ladder} \sim \frac{2}{ 1 + \frac{1}{ n_{\ell} \sqrt{2}  } }
\label{63}
\end{equation}

\noindent
which converges towards  $d=2$ (plane) from below.
  
A less heuristic proposal is to associate 
$f_{n_{\ell}}$ with  a "finite size" effect in the renormalization 
constants $Z_c$ and $Z_{\chi}$ of the ladder. 
Combining  the recent work of 
Chakravarty \cite{Chak}, together with
the results presented above,  we shall propose a 
finite size correction  of  the  spin susceptibility, spin velocity
and spin-stiffness renormalization 
constants of ladders as follows,

\begin{equation}
\begin{array}{rl}  Z_{\chi}(S,n_{\ell})= & 
2 f_{n_{\ell}}  Z_{\chi}(S)  \\
Z_c(S, n_{\ell})= &  Z_c(S)/\sqrt{ 2 f_{n_{\ell}}} \\ 
Z_{\rho_s}(S,n_{\ell}) = &  Z_{\rho_s}(S) \end{array} 
\label{64} 
\end{equation}

\noindent
where $Z_\chi(S), Z_c(S)$ and $Z_{\rho_s}(S)$ are those
of the 2d spin system (i.e. $n_{\ell} = \infty $).
The last eq. in (\ref{64}) follows from the first two thanks to the relation
$Z_{\rho_s} = Z_{\chi} Z^2_c$. We shall give some numerical support
to (\ref{64}) in the next section.

The general formulation we have introduced above,  will  allow us to discuss
certain important issues concerning the ladders.

\subsection*{ Spin Gap of Even Ladders}

Choosing the columnar-block description of the even ladders we deduce from
table 6 and eq.(\ref{14})  
the following expression for the
spin gap,

\begin{equation}
\Delta_{n_{\ell}}^{(1+1)} \sim J S^2 \; n_{\ell} \; 
{\rm exp}\left( - \pi S n_{\ell} \sqrt{f_{n_{\ell}}} \right) 
\label{65}
\end{equation}

\noindent
which  predicts a exponential decay of the gap as a function
on the number of legs \cite{Ger}. This implies in particular
that the spin gaps of the 2, 4 and 6 legged ladders should be related.
Indeed one finds from numerical results of the isotropic ladders

\begin{equation}
\frac{\Delta_2 \Delta_6 }{\Delta_4^2} \sim 1
\label{66}
\end{equation}

\noindent
where we have used the data of  table 7 together the value
$\Delta_2/J =0.504$ \cite{BDRS,WNS}.

In agreement with (\ref{65}),
Chakravarty has recently derived the exponential fall off of the
gap with $n_{\ell}$ using the 2+1 NLSM
\cite{Chak}. In his approach a  spin ladder of
width  $n_{\ell}$  at zero temperature and periodic
boundary conditions along the rungs, is equivalent 
to a Heisenberg plane of infinite extent at a 
finite temperature inversely proportional to $n_{\ell}$.
This allows the use of the 2+1 NLSM results to study  ladder systems.
In particular the expressions for
the spin  gap  and correlation length are given by ( isotropic ladders
$J'=J$),
\cite{Chak}

\begin{eqnarray}
& \Delta^{(2+1)}_{n_{\ell}} = \frac{16 \pi}{{ e}} J S^2 Z_{\rho_S}
{\rm exp} \left( - \frac{\pi S}{\sqrt{2}} \frac{ Z_{\rho_S}}{Z_c}
\frac{ L}{a} \right) \left( 1 - \frac{1}{\pi \sqrt{2} S} 
\frac{Z_c}{ Z_{\rho_S}} \frac{a}{L} \right)^{-1} &
\label{67} \\
& \xi^{(2+1)}_{n_{\ell}} = \frac{{ e}}{ 4 \pi \sqrt{2}} 
\frac{J  Z_c}{ S Z_{\rho_S}}
{\rm exp} \left(  \frac{\pi S}{\sqrt{2}} \frac{ Z_{\rho_S}}{Z_c}
\frac{ L}{a} \right) \left( 1 - \frac{1}{\pi \sqrt{2} S} 
\frac{Z_c}{ Z_{\rho_S}} \frac{a}{L} \right) &
\nonumber
\end{eqnarray}

\noindent
where L and a are the width and lattice spacing of the ladder
($L/a = n_{\ell}$). 
The exponential behaviour of  eqs(\ref{65}) and (\ref{67}) agree
in the classical limit $S \rightarrow \infty$ provided we choose
$Z_c$ and $Z_{\rho_s}$ as the renormalization constants 
$Z_c(S, n_{\ell})$ and $Z_{\rho_s}(S, n_{\ell})$ 
defined in (\ref{64}). This gives further support to the 
finite size correction propose in (\ref{64}).

In tables 7 and 8 we give the  values of the gap and
correlation length of the 4 and 6 legged ladders 
obtained using: 
i) numerical methods (Quantum Monte Carlo \cite{BAT,GBW} 
and DMRG \cite{WNS}),
ii) the NLSM in 2+1 ( \cite{Chak}) and iii) the finite size correction
of the NLSM results of \cite{Chak}. For the two later set of data the values
of the renormalization constant for S=1/2 are choosen  as $Z_c = 1.18, 
Z_{\rho_s}=0.724$ \cite{Singh}.

\begin{center}
\begin{tabular}{|c|c|c|c|c|}
\hline
$n_{\ell}$ & DMRG & QMC & $NLSM(2+1)$ & $NLSM(2+1)$+finite size   \\
\hline
\hline
4 & 0.190 & 0.16 - 0.17 & 0.268 & 0.209  \\
\hline
6 & ? & 0.055 - 0.05 & 0.064 & 0.050 \\
\hline
\end{tabular}
\end{center}
\begin{center}
Table 7: Ladder's spin gap.  
\end{center}

\begin{center}
\begin{tabular}{|c|c|c|c|c|}
\hline
$ n_{\ell} $ & DMRG & QMC & $NLSM(2+1)$ & $NSLM(2+1)+$finite size \\
\hline
\hline
4 & 5-6  &  10.3   &  6.23 & 7.37 \\
\hline
6 &  ?  & $ \sim 30$   & 26.2 & 31.6 \\
\hline
\end{tabular}
\end{center}
\begin{center}
Table 8: Ladder's correlation length.
\end{center}

We observe from tables 7 and 8 that the 
finite size modification  of the Chakravarty formulas (\ref{67})
seems to give a rather good agreement with the numerical results. 
Further work needs to be done to settle this matter.

\subsection*{Limits of Applicability of the Ladder's Map}

We mentioned at the beginning of the construction of  the map 
from the ladder into a unique NSLM field, that it  
would be  valid for the intermediate coupling region ($J'/J \sim
O(1)) $. We shall next explain this point in more detail.

First of all let us consider the weak coupling region $J'/J <<1$.
As shown in eqs.(\ref{37}) the masses of the higher modes of the ladder,
at the linearized level,  are of order $\sqrt{J J'}$. On the other
hand the mass generated non perturbatively is given by (\ref{65}).  
A consistent  truncation of the massive modes then requires that
the mass of these modes should be larger than the mass generated non
perturbatively, which leads to,

\begin{equation}
A e^{- B n_{\ell}} << \; \frac{J'}{J} 
\label{weak}
\end{equation}

\noindent
This equation implies that there
is a lower critical value, $(J'/J)_c$ below which the truncation of
the high energy modes, at least in the way it is done here, is not valid.
In particular,   
in the weak coupling region the spin  gap is approximately proportional to 
$J'$ ( for $n_{\ell}= 2 , \Delta \sim 0.41 J' $)\cite{BDRS,Hata,GBW}. 
This behaviour is not
consistent with (\ref{65}). 
On the other hand eq.(\ref{weak}) 
suggests that the range of applicability of our
model is bigger as $n_{\ell}$ increases.

Let us now consider the strong coupling regime ($J'/J>>1)$, and choose
as an example the two legged ladder. 
For the columnar block, the map (\ref{40}) reads,

\begin{equation}
\begin{array}{cl} S_1(n) = &  \frac{1}{2} {\bf l}(n)  + S (-1)^n  
{\bf \Phi}(n) \\
S_2(n) = & \frac{1}{2} {\bf l}(n)  - S  (-1)^n
{\bf \Phi}(n) \end{array}
\label{map}
\end{equation}

\noindent
which plugged into the ladder Hamiltonian leads, without 
making any approximation, to

\begin{equation}
H_{\rm ladder} = \sum_n \frac{J'}{2} \; {\bf l}(n)^2 +
J \left( \frac{1}{2} {\bf l}(n)\; {\bf l}(n+1) - 2 S^2 
{\bf \Phi}(n) \; {\bf \Phi}(n+1) \right) 
\label{ham}
\end{equation}

If we now apply a gradient expansion in the fields ${\bf l}$ and
${\bf \Phi}$, we obtain the results given in (\ref{43})
(see also table 9), for the case $n_{\ell}=2$. 
It is interesting to compare (\ref{ham}) with the lattice 
NLSM Hamiltonian (\ref{12}). We see that the gradient expansion and
truncation of the ${\bf l}'s$ field is crucial in order that
(\ref{ham}) becomes a NLSM Hamiltonian. In the strong coupling regime,
as can be seen from the mean field analysis of \cite{GRS},
what is pertinent is to drop the term proportional to
${\bf l}(n) {\bf l}(n+1)$ in (\ref{ham}). If we do this approximation
then (\ref{ham}) become the discrete NLSM Hamiltonian with the following
identification of $g $ and $c$,

\begin{equation}
\begin{array}{rl}  c_{\rm latt} = & S \sqrt{2 J J'} \\
                   g_{\rm latt} = & \frac{1}{S} \sqrt{\frac{J'}{2J}}
                  \end{array}
\label{para}
\end{equation}

If we replace these parameters into the gap formula obtained
in the strong coupling regime (\ref{13}) we get

\begin{equation}
\Delta/J' = 1  - \frac{ 4 S^2}{3} \frac{J}{J'} + 0.296 S^4 
\left(  \frac{J}{J'} \right)^2  
\label{sgap}
\end{equation}

If we replace in (\ref{sgap}) $S^2 $ by $S(S+1)$ and we particularize 
to the case  $S= 1/2$, then we get 
the correct behaviour of the gap in perturbation
theory  to order J/J' \cite{RTR},

\begin{equation}
\Delta/J' = 1 - \frac{J}{J'} + \frac{1}{2} 
\left(\frac{J}{J'}\right)^2
+ \frac{1}{4} \left(\frac{J}{J'}\right)^3 + 
O( \left(\frac{J}{J'}\right)^4)
\label{pgap} 
\end{equation}

The previous discussion 
illustrates that in the strong coupling limit one 
has to be careful in making 
gradient expansions of the NLSM fields.
A more detailed analysis is needed to clarify this matter.

\subsection*{ Haldane Phase in the 2-Legged Ladder}

An important question concerning even spin ladders is wether these
systems are in a fundamentally new state or they are in a more
familiar state, as for example the integer spin chains.
This problem has been addressed by various authors arriving at 
different conclusions \cite{ Hida, White}. 
In this section we shall apply the NLSM techniques to clarify
this issue, finding support for  
the view of \cite{White} that the 2 legged ladder is in the same
phase than the $S= 1$ chain \cite{Dago}. 
A more detailed discussion is delayed
to the future.

The $n_{\ell}=2$  ladder can 
be studied using columnar, rectangular and diagonal
blocks. The values of the corresponding NLSM parameters are given in
table 9.

\begin{center}
\begin{tabular}{|c|c|c|c|}
\hline
Block &  $\theta$ & $c$ & $g$   \\
\hline
\hline
Columnar &  0 & $2 J S  \left(1 + \frac{J'}{2 J } \right)^{1/2} $& 
$ \frac{1}{S} \left( 1 + \frac{J'}{2J} \right)^{1/2} $     \\
\hline
Rectangular &  0 & 
$2 \sqrt{2} J S  \left(1 + \frac{J'}{2 J } \right)^{1/2} $ &  
$\frac{1}{S \sqrt{2}} \left( 1 + \frac{J'}{2J} \right)^{1/2} $ \\
\hline
Diagonal &  $4 \pi S $ &$2 J S  \left(1 + \frac{J'}{2 J } \right)$
&$\frac{1}{S}$ \\
\hline
\end{tabular}
\end{center}
\begin{center}
Table 9: NLSM parameters of the 2 legged ladder
\end{center}

From table 9 we see that the parameters obtained using
the diagonal blocks coincide with 
those of an effective  spin chain with spin and
exchange coupling given by,

\begin{equation}
S_{\rm eff} = 2 S,  \;\; J_{\rm eff} = \frac{1}{4} ( 2 J + J')
\label{68}
\end{equation}

This is not an accidental fact.
An alternative way to arrive 
to (\ref{68}) is  to construct an effective model in terms
of the  spin 2S states  formed  out by symmetrization of  
two spin S states located  in diagonal positions
of the ladder ( see  fig. 5) \cite{White}, i.e.

\begin{equation}
{\bf S}_{\rm eff}(n) = {\bf S}_1(n) + {\bf S}_2(n-1) 
\label{69}
\end{equation}

The Hamiltonian governing this effective spins  is given, to
lowest order in perturbation theory by the chain Hamiltonian,

\begin{equation}
H_{\rm eff} = \frac{1}{4} (2 J + J') \sum_n {\bf S}_{\rm eff}(n) 
\; {\bf S}_{\rm eff}(n +1) 
\label{70}
\end{equation}

The NLSM  parameters corresponding to this model (\ref{27}) are 
precisely the ones given in table 9 for the diagonal blocks.
Hence  the 2 legged ladder is mapped into the same
NLSM model as the S=1 chain, which suggests  that both 
systems are in the
same phase. The relationship between 
the "standard" phase of the ladder, given 
by the RVB picture of \cite{WNS},   
and that of the spin 1 chain appears, from the point of view of the NLSM, as
the relation between two different types of blockings, namely
the rectangular and the diagonal ones. This relation is given by,

\begin{equation}
\tilde{{\bf l}} = {\bf l} + S {\bf \Phi}', \;  \tilde{{\bf \Phi}}= {\bf \Phi}
\label{71} 
\end{equation}

\noindent
where $ {\bf l}, {\bf \Phi}$ ( resp.
$ \tilde{ {\bf l}} ,  \tilde{ {\bf \Phi} }) $ 
are the NLSM fields associated to rectangular (resp. diagonal) blocks. 
Eq.(\ref{71}) is a cannonical transformation, identical to (\ref{29}),
which changes $ \theta $ from 0 into $4 \pi S$,
leaving invariant the values of $c$ and $g$. 
This poses a puzzle since the later parameters are different
for rectangular and diagonal blocks, except in the
case $J' = 2J$.  This discrepancy 
must be understood along the lines of our discussion
about the different values of $g$ for 
columnar and rectangular blocks, and very likely  do not affect the conclusion
about the equivalence 
between the 2 legged ladder and the spin 1 chain.

\section*{Spin Ladders with Dimerization}

The main property of uniform spin ladders
is that there are no
phase transitions as one varies the  ratio $J'/J$. It is thus
interesting to investigate the existence of new phases by enlarging the
parameter space of the model. There are plenty of possibilites at hand:
dimerization, frustration, spin deffects, etc. We shall review
here the first one.
There are also many possible types of dimerization in a ladder. We
shall consider below dimerizations only along the legs. In this case
we may distinguish between columnar and staggered dimerizations, 
for which the intraleg coupling constant is given by,

\begin{equation}
J_a(n) = \left\{ \begin{array}{rl} 
                  J ( 1 + (-1)^{n} \gamma) & \;\; ({\rm columnar}) \\ 
                  J ( 1 + (-1)^{n +a} \gamma) & \;\;({\rm staggered})
\end{array} \right. 
\label{72}
\end{equation}

\noindent
The dimerization parameter $\gamma$ will be choosen to vary in the
interval $(1, -1)$ in order not to change the AF character of the legs.
The rung coupling constant $J'$ may be positive or
negative, so  all together 
there are 4 types of models.
 In references \cite{Totsuka} 
were studied the models with columnar dimerization
and ferromagnetic rung coupling. If $J' \rightarrow - \infty $
the dimerized ladder becomes effectively
a dimerized chain and one can apply the results
known for  chains. 

A chain with spin S and alternation parameter
$\gamma$ can be mapped into a NLSM with $\theta$ given 
by \cite{AffLH, AH},

\begin{equation}
\theta = 2 \pi S ( 1 + \gamma)
\label{73}
\end{equation}

The criteria that the NLSM models with
$\theta = \pi $ ( mod $2 \pi$) are masless \cite{Hal,Shan}
implies the existence of 2S critical points \cite{AH}.
Indeed, as 
$\gamma$ varies from -1 to 1, the parameter $\theta$, given in 
(\ref{73}), passes 2S times through $\pi$. 
If S=1/2 there is a single  critical point 
corresponding to the non dimerized chain (i.e. $\gamma=0$).
If S=1 there should exists two critical points for $\gamma_c = \pm 1/2$.
Numerical computations show that there are indeed two critical points
located at $\gamma_c  \sim \pm 0.25 $ \cite{Kato}. 
Hence the NLSM predicts the existence
of these critical points but it
is not precise about their localization. The NLSM prediction
has also been confirmed for S= 3/2 \cite{Yajima} and S=2 \cite{Yamanaka}.

Returning to  ladders with columnar dimerization 
we expect, 
from the  above discussion,  that there should exist
two  critical  lines in the plane ($\gamma, J'/J$).
In the example of the spin 1/2 two legged ladder there should
exist two of these lines  
emanating at $J'/J = -\infty , \gamma_c = \pm 0.25$ and ending at the
origin (i.e. $ J'/J = 0 , \gamma =0$) \cite{Totsuka}.  This critical lines
separate the Haldane phase associated to the strong ferromagnetic
ladder and a dimerized phase of weakly coupled chains.

The staggered dimerization with AF rung couplings 
has been studied in \cite{MSS}. The behaviour of these ladders
is very interesting and it is based on the map  
into the NLSM. The value of the $\theta$ parameter 
of a staggered ladder with AF rung coupling is
given by \cite{MSS},

\begin{equation}
\theta = 2 \pi S n_{\ell} ( 1 + \gamma \; f_{n_{\ell}} ) 
\label{74}
\end{equation}

This formula contains (\ref{73}) as a particular case
(for  $n_{\ell}=1$). Using the 
properties of $f_{n_{\ell}}$ 
one can conjecture from (\ref{74}) the  existence of  $2 S n_{\ell}$ 
critical lines in the plane $(\gamma, J'/J)$ \cite{MSS}. 
In the particular case of the spin 1/2 two legged ladder 
there should exist two critical lines separating also 
the RVB phase of the uniform ladders and a dimerized
phase of weakly couple chains (see figure 6). 
The pahse diagram of the 3 legged ladder (S=1/2) contains 
besides the uniform critical line ( $\gamma =0$) two more
critical lines ending at the walls $|\gamma| =1 $ (see figure 7).

The two cases analysed above ( i.e. columnar/F-rung and staggered/AF-rung )
have a very similar phase diagram which suggests some kind of relationship 
between them \cite{MSS2}. Other types of dimerizations will 
be considered in \cite{MSS2}.

{\bf Acknowledgements} I would like to thank T.M. Rice
for introducing me to the subject of ladders, 
M.A. Martin-Delgado  and  R. Shankar for reading this manuscript
and  E. Dagotto, S. Haas,
B. Frischmuth, S. White, H.J. Schulz and S. Sachdev and J. Dukelski
for  conversations, advises and suggestions.

\newpage

\section*{Figure Captions}

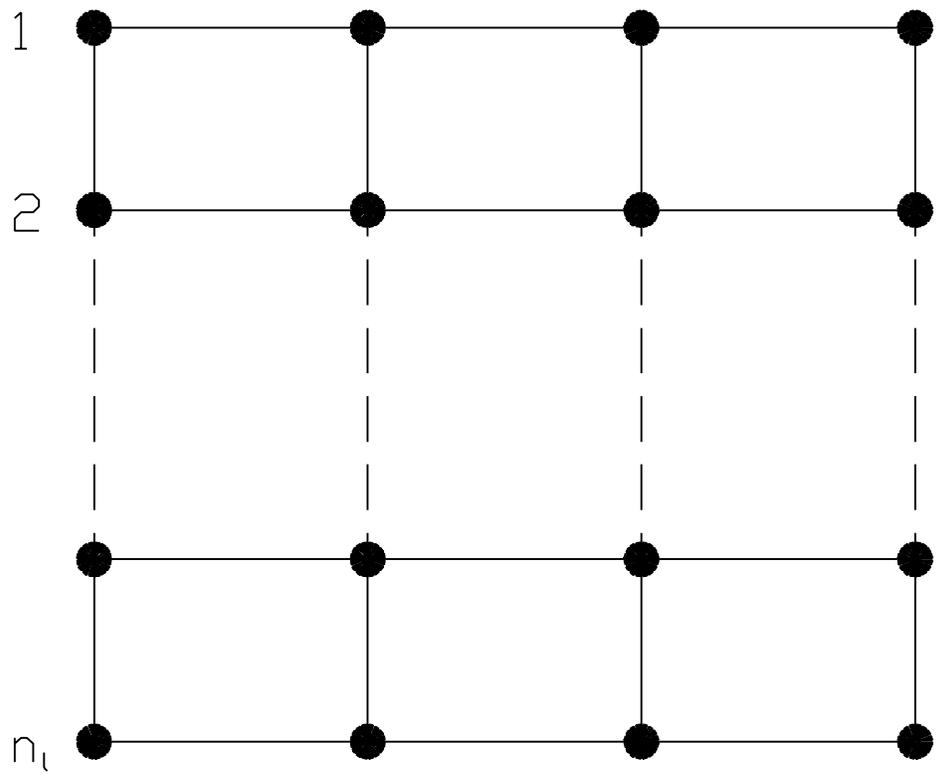
Fig.1.- A generic spin ladder with $n_{\ell}$ legs.

\noindent
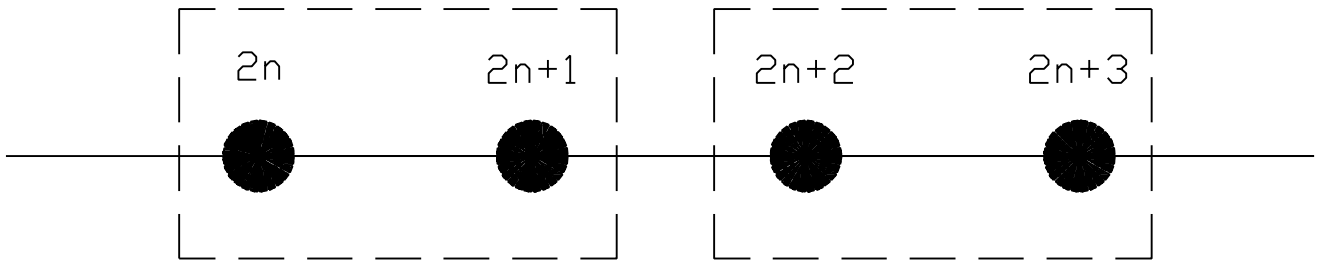
Fig.2.- Blocking of the spin chain needed to define the NLSM
variables out of the spin ones.

\noindent
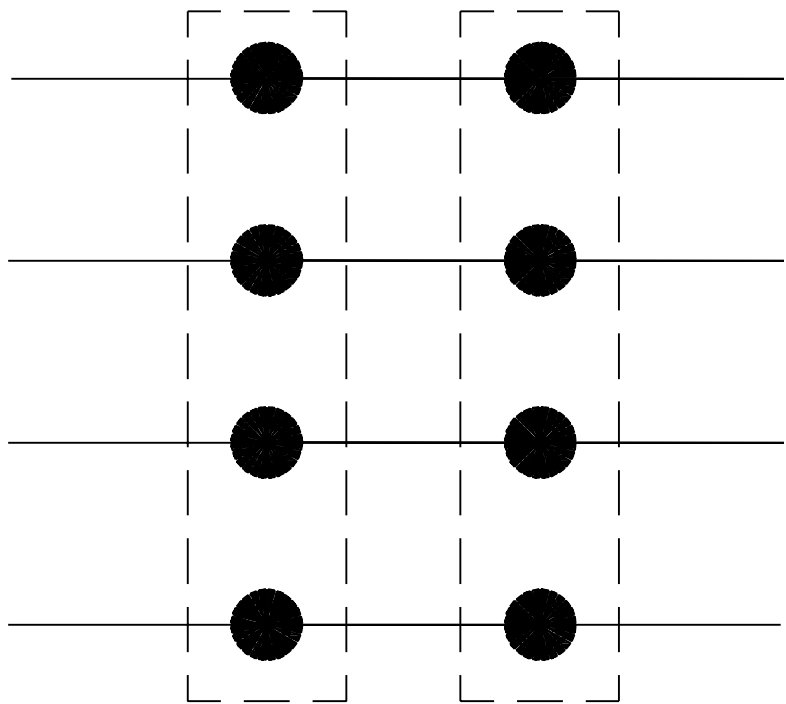
Fig.3.- Columnar Blocking ($n_{\ell}=4$).

\noindent
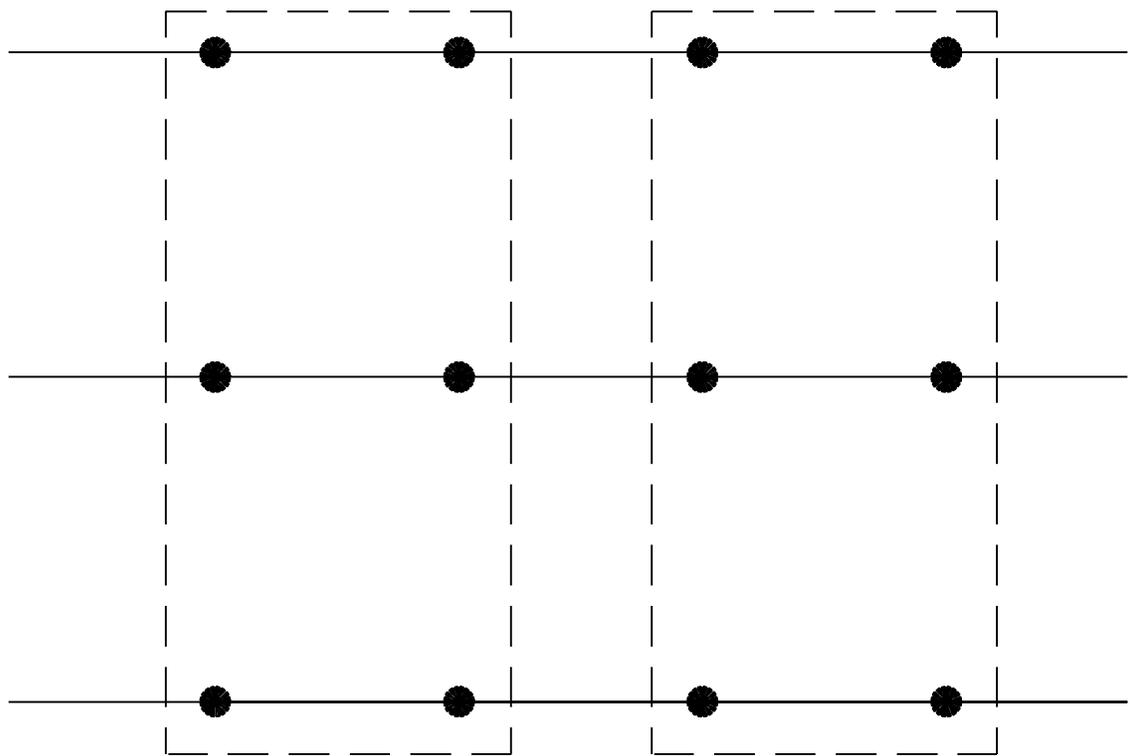
Fig.4.- Rectangular Blocking ($n_{\ell}=3$).

\noindent
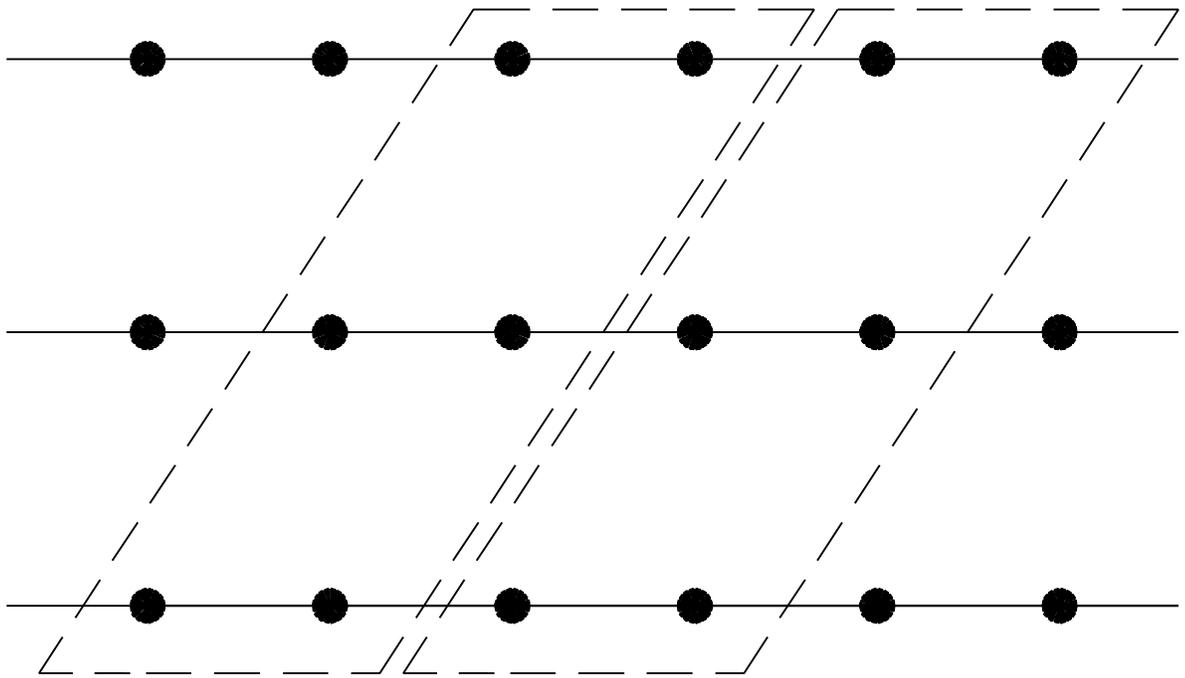
Fig.5.- Diagonal Blocking ($n_{\ell}=3$).

\noindent
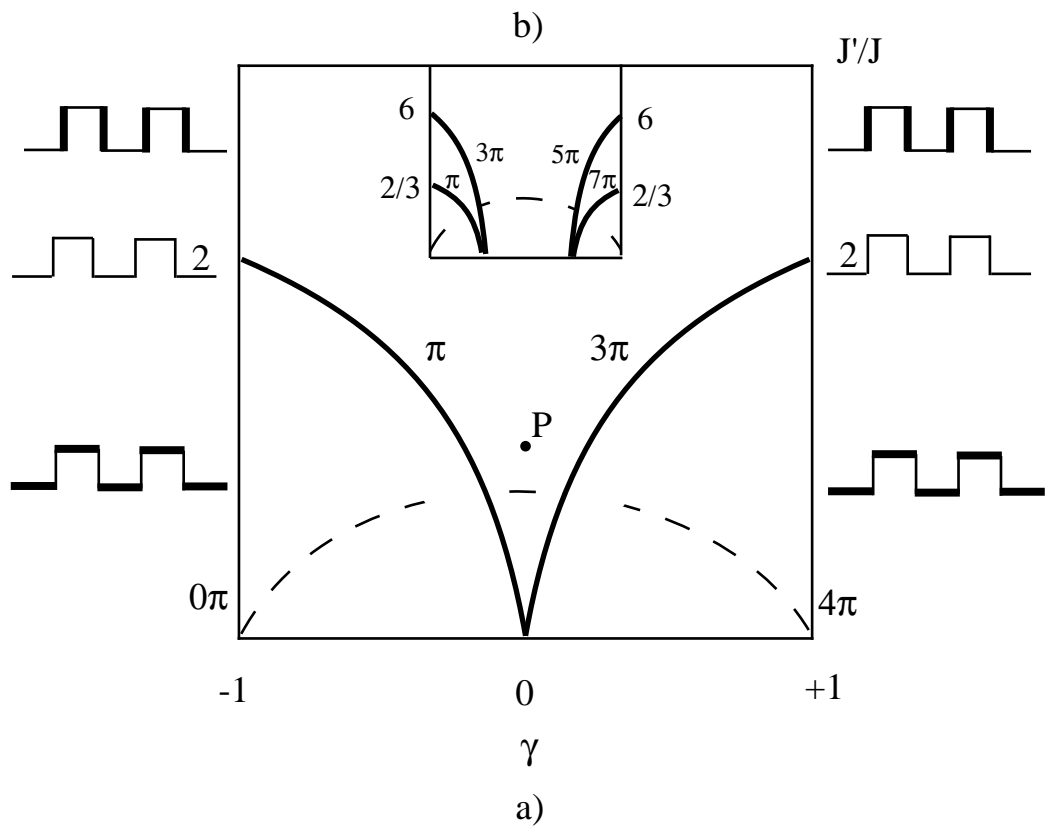
Fig.6.- Phase diagrams of the  2 legged ladder 
with staggered
dimerization for S=1/2 and
S=1 (inset). If $|\gamma| =1 $ the ladders degenerate into
the "snake" chains which are depicted in the margins of the figure.

\noindent
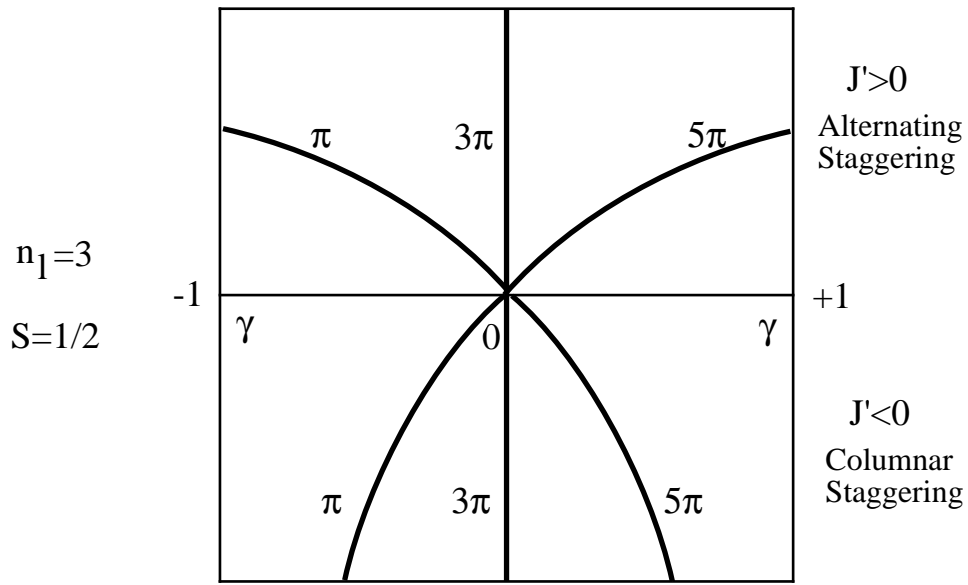
Fig.7.- Phase diagram of the 3 legged ladder with columnar (lower half plane)
and staggered dimerizations (upper half plane). 
Observe the existence of 3 critical lines that emerge from the origin.

\end{document}